\documentclass[fleqn,usenatbib]{mnras}
\usepackage{newtxtext,newtxmath}
\usepackage[T1]{fontenc}
\DeclareRobustCommand{\VAN}[3]{#2}
\let\VANthebibliography\thebibliography
\def\thebibliography{\DeclareRobustCommand{\VAN}[3]{##3}\VANthebibliography}
\usepackage{graphicx}	
\usepackage{amsmath}	
\usepackage{multirow}

\usepackage{ulem}
\usepackage{color,xcolor}
\usepackage{float}

\title[Bayesian analysis using RMF model]{Bayesian analysis of a relativistic hadronic model constrained by recent astrophysical observations}

\author[B. A. de Moura S., C. H. Lenzi, O. Louren\c{c}o, M. Dutra]{
Bruno A. de Moura S.,$^{1}$ \thanks{Contact e-mail: bmoura@ita.br}
C\'esar H. Lenzi,$^{1}$
Odilon Louren\c{c}o,$^{1,2}$
and Mariana Dutra$^{1,2}$ \thanks{Contact e-mail: marianad@ita.br}
\\
$^{1}$ Departamento de F\'isica, Instituto Tecnol\'ogico de Aeron\'autica, DCTA, 12228-900, S\~ao Jos\'e dos Campos, SP, Brazil\\
$^{2}$ Universit\'e de Lyon, Universit\'e Claude Bernard Lyon 1, CNRS/IN2P3, IP2I Lyon, UMR 5822, F-69622, Villeurbanne, France\\
}

\pubyear{2023}

\begin{document}
\label{firstpage}
\pagerange{\pageref{firstpage}--\pageref{lastpage}}
\maketitle

\begin{abstract}
We use Bayesian analysis in order to constrain the equation of state for nuclear matter from astrophysical data related to the recent measurements from the NICER mission, LIGO/Virgo collaboration, and probability distributions of mass and radius from other 12 sources, including thermonuclear busters, and quiescent low-mass X-ray binaries. For this purpose, we base our study on a relativistic hadronic mean field model including an $\omega-\rho$ interaction. Our results indicate optimal ranges for some bulk parameters at the saturation density, namely, effective mass, incompressibility, and symmetry energy slope ($L_0$). For instance, we find $L_0 = 50.79^{+15.16}_{-9.24}$~MeV (Case 1) and $L_0 = 75.06^{+8.43}_{-4.43}$~MeV (Case 2) in a $68\%$ confidence interval for the 2 cases analyzed (different input ranges for $L_0$ related to the PREX-II data). The respective parametrizations are in agreement with important nuclear matter constraints, as well as observational neutron star data, such as the dimensionless tidal deformability of the GW170817 event. From the mass-radius curves obtained from these best parametrizations, we also find the ranges of $11.97~\mbox{km}\leqslant R_{1.4}\leqslant 12.73~\mbox{km}$ (Case 1) and $12.34~\mbox{km}\leqslant R_{1.4}\leqslant 13.06~\mbox{km}$ (Case 2) for the radius of the $1.4M_\odot$ neutron star. 
\end{abstract}


\section{Introduction} 

Neutron stars (NS) are some of the most fascinating and mysterious objects in the universe. These ultra-dense celestial bodies, formed by the collapse of massive stars, offer a unique opportunity to study the properties of nuclear matter under extreme conditions, since their core contains matter up to around ten times the saturation density ($\rho_0 \simeq 2.5 \times 10^{14}$~g/cm$^3\simeq 0.16$~fm$^{-3}$). This environment is a suitable framework in order to test the many types of hadronic relativistic and nonrelativistic models, along with its huge amount of related parametrizations. In that direction, it is possible to select, or at least distinguish, among different models/parametrizations able to reproduce known properties of this astrophysical compact objects~\citep{oertel,lattimer-review,debora-universe}. On the terrestrial side, on the other hand, studies on heavy ion collisions provide some constraints on the equation of state (EoS) for symmetric nuclear matter (SNM) at densities around $\rho_0$~\citep{pawel2002,FUCHS20061}. Other studies based on chiral effective field theory~\citep{Hebeler_2013,Drischler_2016,Kruger_2013} also give some constraints on the EoS for pure neutron matter (PNM).

Recent measurements of NS properties such as masses, radii, and tidal deformability, have opened the door to a new era of investigation on the nature of these objects by establishing new constraints to be satisfied by different models. Such observations include the detection performed by the LIGO and Virgo Collaboration~(LVC) of gravitational waves emitted during the so-called GW170817 event, in which the merge of two NS was registered~\citep{Abbott_2017,Abbott2018,Abbott:prx}, and data provided by the NASA’s Neutron Star Interior Composition Explorer (NICER) mission concerning the X-ray detection coming from the proximity of the pulsars PSR~J0030+0451~\citep{Riley_2019,Miller_2019} and PSR~J0740+6620~\citep{Miller_2021,Riley_2021}. By using the Tolman-Oppenheimer-Volkoff equations (TOV)~\citep{tov39,tov39a}, it is possible to create a link between a specific EoS and a unique mass-radius diagram, a plot in which some of these observational data can be represented. Therefore, different EoSs can be tested against the updated astrophysical constraints. In this paper, we follow this procedure by means of a Bayesian analysis, namely, a powerful computational tool that can be used to combine observational/experimental data with theoretical models to test and constraint them as much as possible, in order to search for suitable EoSs capable of reproducing the aforementioned stellar matter requirements~\citep{Silvia2020,Salinas2023,Zhu2022}. In this sense, works doing EoS inference with Bayesian methods from the LIGO-Virgo collaboration has also been made \citep{Abbott_2019b,Abbott_2020}

We intend to find parametrizations of a relativistic mean-field~(RMF) hadronic finite-range model based on theoretical grounds of the so-called quantum hadrodynamics (QHD). It considers protons and neutrons interacting with each other through Yukawa-type potentials~\citep{yukawa1935interaction}, in which the nucleon-nucleon interaction range is regulated by the masses of the exchanged mesons. Furthermore, the attractive and repulsive strengths of this interaction are controlled by the free parameters of the model. We use a version of the RMF model that takes into account cubic and quartic self-interactions in the field $\sigma$ (attractive, meson $\sigma$), along with a crossed interaction between the fields $\omega^\mu$ (repulsive, meson $\omega$) and $\vec{\rho}_\mu$. The latter represents the isovector meson $\rho$ responsible for the asymmetry of the system (different number of protons and neutrons). The procedure adopted here is to obtain, from the Bayesian inference, the suitable ranges for some nuclear matter bulk parameters and generate, from these quantities, the optimal parametrizations of this RMF model. As input for the method, we furnish Likelihood functions extracted from 16 different sources of masses and radii from thermonuclear busters~\cite{ozel2016dense}, from quiescent low-mass X-ray binaries~\cite{ozel2016dense}, from NICER, and LVC. We show that these ``best'' parametrizations determined from this approach are able to describe symmetric nuclear matter, pure neutron matter, and properties of the stellar matter system as well (radius of the $1.4M_\odot$ neutron star, and tidal deformability, for instance). 

In comparison to the study performed in~\cite{Silvia2020}, our work presents the following novelties: (i) we use a Lagrangian density for the relativistic model with a $\omega-\rho$ interaction included. This new term gives us the freedom of choosing one more isovector bulk parameter to be used in the Bayesian analysis. In our case, in addition to the symmetry energy slope, we also take the symmetry energy at $2/3$ of the saturation density; (ii) the range used for the former is based on the study coming from the analysis of the PREX-II results for the $^{208}\rm Pb$ neutron skin thickness~\citep{PREX-ESSICK,PREX-PRL126,PREX-PRR4-L022054,PREX-PRL126-172503}; (iii) we used as one of the sources of the Bayesian analysis the recent observational data provided by the NICER mission with regard to the pulsar PSR J0704+6620~\citep{Riley_2021}.

The paper is organized as follows. In Section~\ref{theo} we present the theoretical framework of the RMF model used as well as the main equations defining the neutron star matter properties. In Section~\ref{sec:Bayes} we furnish the details of the Bayesian method and the inputs used. Section~\ref{sec:Results} is dedicated to the results of our study, and, finally, summary and final remarks are shown in Section~\ref{sec:Final}.

\section{Theoretical framework} 
\label{theo}

\subsection{Relativistic mean-field model and bulk parameters}
\label{sec:Equationofstate}

The RMF model used here is described by the following Lagrangian density~\citep{Dutra2014,LI2008113}
\begin{equation}
\begin{aligned}
\mathcal{L} = &\Bar{\Psi}(i\gamma^{\mu}\partial_{\mu}-M_{\mbox{\tiny nuc}})\Psi - g_{\sigma}\sigma\Bar{\Psi}\Psi -g_{\omega}\Bar{\Psi}\gamma^{\mu}\omega_{\mu}\Psi\\ &-\frac{1}{2}g_{\rho}\Bar{\Psi}\gamma^{\mu}\Vec{\rho}_{\mu}\Vec{\tau}\Psi + \frac{1}{2}(\partial^{\mu}\sigma\partial_{\mu}\sigma-m_{\sigma}\sigma^{2})\\
&-\frac{A}{3}\sigma^{3}-\frac{B}{4}\sigma^{4}-\frac{1}{4}F^{\mu\nu}F_{\mu\nu}-\frac{1}{4}\Vec{B}^{\mu\nu}\Vec{B}_{\mu\nu}\\
&+\frac{1}{2}m_{\rho}^{2}\Vec{\rho}_{\mu}\Vec{\rho}^{\mu}+ \frac{1}{2}m_{\omega}^{2}\omega_{\mu}\omega^{\mu}+\frac{1}{2}\alpha g_{\omega}^{2}g_{\rho}^{2}\omega_{\mu}\omega^{\mu}\Vec{\rho}_{\mu}\Vec{\rho}^{\mu},
\label{Main_Lagrangian}
\end{aligned}
\end{equation}
where the nucleon field is represented by the Dirac spinor $\Psi$. The scalar, vector, and isovector fields $\sigma$, $\omega^\mu$, and $\vec{\rho}_\mu$ represent the mesons $\sigma$, $\omega$, and $\rho$, respectively. The antisymmetric tensors are given by $F_{\mu\nu} = \partial_\nu\omega_\mu - \partial_\mu\omega_\nu$, and $\vec{B}_{\mu\nu} = \partial_\nu\vec{\rho}_\mu - \partial_\mu\vec{\rho}_\nu$. The nucleon mass is $M_{\mbox{\tiny nuc}}=939$~MeV and the mesons masses are $m_\sigma$, $m_\omega$, and $m_\rho$. By using the mean-field approximation along with the Euler-Lagrange equations, it is possible to construct the energy-momentum tensor $T^{\mu\nu}$. From this quantity, one obtains pressure, and energy density, in this case, given by 
\begin{equation}
\begin{split}
P&=\frac{\left<T_{ii}\right>}{3} = -\frac{1}{2}m_{\sigma}^2\sigma^2 - \frac{A}{3}\sigma^3 - \frac{B}{4}\sigma^4 +\frac{1}{2}m_{\omega}^2\omega_0^2 + \frac{1}{2}m_{\rho}^2\Bar{\rho_0}^2\\ 
&+\frac{1}{2}\alpha g_{\omega}^{2}g_{\rho}^{2}\omega_{0}^{2}\bar{\rho_{0}}^{2} + P_{kin}^{n}+P_{kin}^{p},
\end{split}
\label{pressure}
\end{equation}
and
\begin{equation}
\begin{split}
\varepsilon &=\left<T_{00}\right>= \frac{1}{2}m_{\sigma}^{2}\sigma^{2} + \frac{A}{3}\sigma^{3}+\frac{B}{4}\sigma^{4}-\frac{1}{2}m_{\omega}^{2}\omega_{0}^{2}-\frac{1}{2}m_{\rho}^{2}\Bar{\rho_{0}}^{2}\\
&+g_{\omega}\omega_{0}\rho + \frac{g_{\rho}}{2}\Bar{\rho_{0}}\rho_3-\frac{1}{2}\alpha g_{\omega}^{2}g_{\rho}^{2}\omega_{0}^{2}\bar{\rho_{0}}^{2}+\varepsilon_{kin}^{n}+\varepsilon_{kin}^{p},
\label{de}
\end{split}
\end{equation}
respectively, with the kinetic parts written as 
\begin{equation}
P_{kin}^{p,n}=\frac{\gamma}{6\pi^{2}}\int_{0}^{k_{F_{p,n}}}\hspace{-0.3cm}
\frac{dk\,k^4}{(k^2+M^{*2})^{1/2}}
\end{equation}
and
\begin{equation}
\varepsilon_{kin}^{p,n}=\frac{\gamma}{2\pi^{2}}\int_{0}^{k_{F_{p,n}}}
\hspace{-0.3cm}dk\,k^2(k^2+M^{*2})^{1/2}.
\end{equation}
$k_{F_{p,n}}$ is the nucleon Fermi momentum, and $\gamma$ is the degeneracy factor. This kind of hadronic model gives rise to a modification in the nucleon rest mass. In this case, the effective Dirac mass~\citep{baoanli2007,typel2018} reads
\begin{equation}
M^* = M_{\mbox{\tiny nuc}} - g_\sigma \sigma.
\end{equation}
For the sake of convenience, we also define, from this quantity, the ratio $m_0^*=M^*(\rho_0)/M_{\mbox{\tiny nuc}}$, where $\rho_0$ is the saturation density. All the remaining thermodynamical quantities can be obtained through Eqs.~\eqref{pressure} and~\eqref{de}. For instance, the nuclear matter incompressibility is found by taking $K=9(\partial P/\partial\rho)$, with the respective value at the saturation density being $K_0\equiv K(\rho=\rho_0)$, and the binding energy of the system is obtained as $E_0\equiv \varepsilon(\rho_0)/\rho_0-M$. For the isovector part, the symmetry energy for instance is $\mathcal{S}(\rho) =
\frac{1}{8}\frac{\partial^{2}(\varepsilon/\rho)}{\partial y^{2}}\Big|_{y=\frac{1}{2}}$, with $y$ being the proton fraction of the system. 

In general, these bulk parameters can be used in order to write the density dependence of the energy density for symmetric nuclear matter~\citep{PhysRevC.85.035201,PhysRevC.80.045806}, namely, 
\begin{equation}
E_{SNM} = E_0 + \frac{1}{2} K_0 x^2 + O(x^3),
\end{equation}
with $x=(\rho-\rho_0)/3\rho_0$. Similarly, the density dependence of the symmetry energy in SNM can be written as~\citep{PhysRevLett.102.122502}
\begin{equation}
\mathcal{S} = J + L_0x + O(x^2),
\end{equation}
where the quantities $J$ and $L_0$ are given, respectively, by
\begin{equation}
J = \mathcal{S}(\rho_0) \quad \mbox{and} \quad 
L_0 = \rho_0\left(\frac{\partial\mathcal{S}}{\partial\rho} \right)_{\rho=\rho_0}.
\end{equation}

It is worth mentioning that many hadronic models exhibit a strong correlation between the symmetry energy at the saturation density ($J$) and its slope ($L_0$)~\citep{Gandolfi_2012,Hebeler_2013,PhysRevC.92.015210,PhysRevLett.125.202702,universe7060182}. In order to reproduce this feature, we decide to use as input in our calculations the value of the symmetry energy at $\rho_{\mbox{\tiny cross}}=2\rho_0/3$, namely, $\mathcal{S}_{2/3}\equiv \mathcal{S}(\rho_{\mbox{\tiny cross}})$. It was mathematically shown in~\cite{PhysRevC.92.015210} that if different parametrizations of a specific bulk parameter $\mathcal{A}(\rho)$ cross each other when plotted as a function of $\rho$, then the bulk parameter defined as the derivative of $\mathcal{A}$ with respect to the density, namely, $\mathcal{B}(\rho)$, can present a linear dependence with $\mathcal{A}$ at the saturation point, i.e., $\mathcal{B}$ can correlates with $\mathcal{A}$ as $\mathcal{B}(\rho_0)=a + b\mathcal{A}(\rho_0)$. By using the notation of~\cite{PhysRevC.92.015210}, we named the density at which the parametrizations of $\mathcal{A}$ cross each other as the crossing density. Notice that since $L(\rho)$ is defined as a derivative of $\mathcal{S}(\rho)$, then the study performed in~\cite{PhysRevC.92.015210} can be used to impose the correlation between $L_0$ and $J$. In this case, we decided to use the crossing density given by $\rho_{\mbox{\tiny cross}}=2\rho_0/3$ as suggested by~\cite{Khan_2013}. It is worth emphasizing that $\mathcal{S}_{2/3}$ and $L_0$ are the only isovector quantities used as input in the Bayesian analysis, differently from the case of~\cite{PhysRevLett.125.202702}, that uses the correlated bulk parameters $J$ and $L_0$ also quantifying the related uncertainties.

As a remark, we mention that, in principle, hyperons would be included in the model. However, the inclusion of such particles in the system softens the equations of state. As a consequence, the model does not produce very massive stars in comparison with the case in which nucleons are the only degrees of freedom (the hyperon puzzle). Moreover, hyperons interactions are not completely determined and some assumptions need to be taken, which makes inevitable the proliferation of more coupling constants to be fixed in the model. In this work, we prefer to avoid such issues by keeping, for the sake of simplicity, only nucleons and mesons on the hadronic side of the stellar matter system.

Here we use some particular ranges for the bulk parameters $\rho_0$, $m_0^*$, $E_0$, $K_0$, $\mathcal{S}_{2/3}$, $L_0$, and from them, we determine the six coupling constants of the model, namely, $g_\sigma$, $g_\omega$, $g_\rho$, $A$, $B$ and $\alpha$. These parametrizations are used as input for the Bayesian analysis. The ``best'' parametrizations obtained as the output of this procedure are shown to satisfy the observational constraints on the mass-radius diagrams, as well as the tidal deformability of the GW170817 event. We show these results and the aforementioned ranges in Sections~\ref{sec:Bayes} and~\ref{sec:Results}.

\subsection{Neutron star matter and tidal deformability} \label{sec:TOV}

For the calculation of the mass-radius diagrams, it is needed to solve the Tolman-Oppenheimer-Volkoff (TOV) equations given by~\citep{tov39,tov39a}, with the total pressure and energy density are given by,
\begin{eqnarray} 
p = P + \frac{\mu_e^4}{12\pi^2}
+ \frac{1}{3\pi^2}\int_0^{\sqrt{\mu_\mu^2-m^2_\mu}}\hspace{-0.3cm}\frac{dk\,k^4}{(k^2+m_\mu^2)^{1/2 }},\qquad\,
\label{totalp}
\end{eqnarray}
and 
\begin{eqnarray}
\mathcal{\epsilon} = \varepsilon + \frac{\mu_e^4}{4\pi^2}
+ \frac{1}{\pi^2}\int_0^{\sqrt{\mu_\mu^2-m^2_\mu}}\hspace{-0.3cm}dk\,k^2(k^2+m_\mu^2)^{1/2},\qquad
\label{totaled}
\end{eqnarray}
respectively. For the leptonic part of the system, we include muons ($\mu$), with $m_\mu=105.7$~MeV and massless electrons ($e$). For each baryonic density, one needs to consider charge neutrality condition and chemical equilibrium $\mu_n - \mu_p = \mu_e$ and $\rho_p - \rho_e = \rho_\mu$, where $\mu_{p,n}$ are the protons/neutrons chemical potentials. For the neutron star crust, we use the Baym-Pethick-Sutherland EoS~\citep{bps} for a density range of $6.29\times10^{-12} \leqslant\rho\leqslant 8.91\times10^{-3}$~fm$^{-3}$.


In binary neutron star systems, such as that in which the LVC has detected for the first time the GW produced by the coalescence of two NS~\citep{Abbott_2017,Abbott2018}, the induced quadrupole momentum in one star, due to the static tidal field generated by the companion one, is related to the so-called tidal deformability parameter $\lambda$. The dimensionless version of such quantity reads
\begin{equation}
\Lambda = \frac{2}{3}\frac{k_2}{C^5},
\label{tidal}
\end{equation}
given in terms of $C=M/R$ (compactness parameter) and $k_2$ (second tidal Love number). The latter is calculated through
\begin{eqnarray}
&k_2& = \frac{8C^5}{5}(1-2C)^2[2+2C(y_R-1)-y_R]\nonumber\\
&\times&\Big\{2C [6-3y_R+3C(5y_R-8)] \nonumber\\
&+& 4C^3[13-11y_R+C(3y_R-2) + 2C^2(1+y_R)]\nonumber\\
&+& 3(1-2C)^2[2-y_R+2C(y_R-1)]{\rm ln}(1-2C)\Big\}^{-1},\qquad
\label{k2}
\end{eqnarray}
with $y_R\equiv y(R)$. The function $y(r)$ is found by solving the differential equation, coupled to the TOV ones, given by
\begin{equation}
r(dy/dr) + y^2 + yF(r) + r^2Q(r)=0, 
\end{equation}
with
\begin{eqnarray}
F(r) &=& \frac{1 - 4\pi r^2[\epsilon(r) - p(r)]}{1-2m(r)/r} , 
\\
Q(r)&=&\frac{4\pi}{g(r)}\left[5\epsilon(r) + 9p(r) + 
\frac{\epsilon(r)+p(r)}{\partial p(r)/\partial\varepsilon(r)}- \frac{6}{4\pi r^2}\right]
\nonumber\\ 
&-& 4\left[ \frac{m(r)+4\pi r^3 p(r)}{r^2(1-2m(r)/r)} \right]^2.
\label{qr}
\end{eqnarray}
More details regarding the derivation of these equations can be seen in ~\cite{Hinderer_2010,Postnikov_2010,Damour_2010,Binnington_2009,Hinderer_2008}.

\section{Bayesian Method}
\label{sec:Bayes}

The method used to apply the mass-radius relation constraints to the equation of state of the RMF model discussed before is presented in this Section, namely, the Bayesian analysis. This analysis is a way to test and constrain the models. In this approach, the main idea is the construction of a posterior probability via Bayes' Theorem~\citep{sivia2006data,gelman2003bayesian}, that reads
\begin{eqnarray}
    P(\theta|D,\mathcal{M})=\frac{P(D|\theta,\mathcal{M})P(\theta,\mathcal{M})}{P(D,\mathcal{M})}.
\label{BayesEquation}
\end{eqnarray}
It allows the calculation of the probability distribution of a set of parameters $\theta$, given the data $D$ for a given model $\mathcal{M}$. The distribution $P(\theta|D,\mathcal{M})$ is called {probability distribution function (PDF). It is the result of the joining of previous information, given by the prior probability $P(\theta,\mathcal{M})$ that reflects the knowledge before the data is considered, with the Likelihood function $P(D|\theta,\mathcal{M})$ that is the statistical data from some probabilistic models. The probability of the observable data $D$, $P(D,\mathcal{M})$, is known as the evidence and it can be treated as a normalization factor since it does not depend on the parameters $\theta$~\citep{Steiner_2010,Silvia2020}. 

As stated in Section~\ref{sec:Equationofstate}, we determine our six coupling constants of the model from the six bulk parameters $m_0^*$, $E_0$, $K_0$, $\rho_0$, $\mathcal{S}_{2/3}$, and $L_0$. Based on different theoretical studies \citep{PRC101,jerome_2018,PRC93BaoJunCai}, we decide to fix $\rho_0$ and $E_0$ because it is well established in the literature that their values are around $0.15$~fm$^{-3}$ and $E_0$ at $-16.0$~MeV, respectively, with very small error bars, as one can see for instance in~\cite{jerome_2018}, in which the ranges for such quantities are given by $\rho_0=(0.155\pm0.005)$~fm$^{-3}$ and $E_0=(-15.8\pm0.3)$~MeV. The remaining bulk parameters constitute our set $\theta$ and will be inferred by the Bayes' Theorem, Eq.~\eqref{BayesEquation}. Different studies were performed in order to determine ranges for these bulk parameters, and we follow some of them to give input for the Bayesian analysis. For $m_0^*$, we chose the range found in~\cite{PRC101}, namely, $0.60\leqslant m_0^*\leqslant 0.88$. The range used for the incompressibility is $220~\mbox{MeV}\leqslant K_0\leqslant 260$~MeV~\citep{GARG201855}. In the case of $\mathcal{S}_{2/3}$, we take the range $22.45~\mbox{MeV}\leqslant \mathcal{S}_{2/3}\leqslant 26.04~\mbox{MeV}$, obtained for the nonlinear RMF models found in~\cite{PRC93_Dutra} (excluding the density-dependent ones), see Table I of this reference. For the symmetry energy slope, we use the findings of recent studies that calculate $L_0$ based on the results for the neutron-skin thickness of $^{208}$Pb experiment (PREX-II), $L_0=53^{+14}_{-15}$~MeV \citep{PREX-ESSICK}, $L_0 = (106\pm37)$~MeV \citep{PREX-PRL126-172503}, $42<L_0<117$~MeV \citep{PREX-PRL126}, $L_0 = (83.1\pm24.7$)~MeV \citep{PREX-PRR4-L022054}. As we also intend to investigate how different ranges of $L_0$ affect the PDF, we decide to split our priors into two cases, namely,
\begin{itemize}
    \item {\bf Case 1}: range determined by the lowest and the highest values of $L_0$:
    \begin{eqnarray}
    38~\mbox{MeV}\leqslant L_0 \leqslant 143~\mbox{MeV}
    \end{eqnarray}
    
    \item {\bf Case 2}: range determined by the the intersection of the $L_0$ values from \cite{PREX-PRL126-172503,PREX-PRL126,PREX-PRR4-L022054}:
    \begin{eqnarray}
    69~\mbox{MeV}\leqslant L_0 \leqslant 107.8~\mbox{MeV}.
    \end{eqnarray}    
\end{itemize}

In principle, different priors may alter the posteriors~\citep{Steiner2016}. In order to verify this hypothesis, we calculate the latter by using Uniform and Gaussian prior distributions. In the case of the Uniform distribution, the global ranges for the bulk parameters presented before are used (minima and maxima values). With regard to the Gaussian distribution, we assign the respective mean value, $\Delta=(Max + Min)/2$, and deviation, $\sigma=(Max-Min)/4$, for each quantity. All the numbers used as input for the different priors are presented in Table~\ref{table:prior_ranges}.
\begin{table}
\centering
\caption{Ranges used in the priors. For saturation density and binding energy, we fix $\rho_0 = 0.15~$fm$^{-3}$ and $E_0=-16$~MeV, respectively.}
\label{table:prior_ranges}
\begin{tabular}{ c c c c c c c } 
\hline
Prior & Range & $m_0^*$ & $K_0$ & $\mathcal{S}_{2/3} $ & \multicolumn{2}{c}{$L_0$}\\
      &       &         &  (MeV)  &  (MeV)  &  \multicolumn{2}{c}{(MeV)} \\
      &       &         &       &                        &   Case 1  &  Case 2 \\
\hline
\multirow{2}{*}{Uniform} & Min & 0.60 & 220.0 & 22.45 & 38.0 & 69.0 \\
& Max & 0.88 & 260.0 & 26.04 & 143.0 & 107.8 \\
\hline
\multirow{2}{*}{Gaussian} & $\Delta$ & 0.74 & 240.0 & 24.25 & 90.5 & 88.4 \\ 
& $\sigma$ & 0.07 & 10.0 & 0.90 & 26.25 & 9.7 \\
\hline
\end{tabular}
\end{table}

It is important to emphasize here that in order to calculate the posteriors, we adopted a method widely used in the literature and detailed, for instance, in~\cite{Steiner_2010,ozel2016dense,Raithel_2017,Silvia2020}. In this case, the likelihood is the probability of generating $N$ mass-radius observations given a particular set of empirical parameters. In our case, given an EoS, we calculate the probability of realization of the pair $D =$~($M$,$R$) of a specific source through the following procedure: (i) given a set of empirical bulk parameters, we calculate the respective EoS and, from that, the corresponding mass-radius diagram. If the EoS produces $M_{{\rm max}}< 2 M_\odot$, the particular parameters set is disregarded and another one is taken. (ii) If the condition $M_{{\rm max}}\geq 2 M_\odot$ is satisfied, the probability of each star configuration of the mass-radius diagram is evaluated by using the $M-R$ distribution of the source. Then, we have 
\begin{eqnarray}
P(D|m_0^*, K_0,\mathcal{S}_{2/3},L_0,\mathcal{M}) =\prod_{i=1}^{n=16} P_i(D_i|m_0^*, K_0,\mathcal{S}_{2/3},L_0,\mathcal{M}),
\end{eqnarray}
where
\begin{eqnarray}
P_i(D_i|m_0^*, K_0,\mathcal{S}_{2/3},L_0,\mathcal{M}) = P_{max,i}(D_i|m_0^*, K_0,\mathcal{S}_{2/3},L_0,\mathcal{M}),
\end{eqnarray}
where $P_{max,i}(D_i|m_0^*, K_0,\mathcal{S}_{2/3},L_0,\mathcal{M})$ is maximum probability associated with the parameter set.

The observational data are constituted of 16 sources, which are posterior probability distributions from masses and radii. 11 of them were extracted from \cite{ozel2016dense}, where 6 are thermonuclear busters called 4U 1820–30, SAX J1748.9–2021, EXO 1745–248, KS 1731–260, 4U 1724–207, and 4U 1608 –52. The data of these sources are calculated by using their apparent angular diameters, distance, and touchdown fluxes. The 5 other sources, out of the 11 ones, were obtained from the following Quiescent Low-Mass X-ray Binaries (qLMXBs): M13, M30, NGC 6304, NGC 6397, and $\omega$~Cen, whose mass-radius posteriors are calculated from the spectroscopic analysis of the thermal radiation emission. All the explanation about these $11$ sources, as well as the $M-R$ data distributions, can be found in~\cite{ozel-data}.

Regarding the 5 sources remaining, out of the 16 ones, one of them comes from the neutron star X-ray analysis in 4U 1702–429, see~\cite{Natalia2017}, in which the authors calculated $M = 1.9 \pm 0.3$~M$_{\odot}$ and $R = 12.4 \pm 0.4$~km. From the LVC~\citep{Abbott2018} we use the 2 components coming from the analysis of the gravitational wave event GW170817, namely, the posterior mass-radius insensitive to EoSs~\citep{samples-ligo}. The last 2 sources come from mass-radius measurements obtained from the data provided by the NICER mission, namely, one for pulsar PSR J0030+0451, calculated through models for the pulsation profile~\citep{Riley_2019}, and another for pulsar PSR~J0740+6620, calculated by using X-ray observations and rotating hot region patterns~\citep{Riley_2021}. In order to mimic the $M-R$ distributions for the 4U 1702–429 and PSR J0030+0451 observables, we use the same description and parameters proposed in~\cite{Natalia2017,Silvia2020}, namely, the bivariate Gaussian distribution.

Finally, the numerator of the PDF in Eq.~\eqref{BayesEquation} is calculated through the Markov Chain Monte Carlo (MCMC). For this purpose, we use the \textit{emcee} python package presented in~\cite{Foreman_Mackey_2013}. However, as our study considers 2 different priors, as well as 2 different sets of parameters $L_0$, it is necessary to calculate the evidence $P(D,\mathcal{M})$. Such a quantity was obtained through a Monte Carlo integration over the posterior.

\section{Results} 
\label{sec:Results}

\subsection{Bayesian analysis}

Following the procedure detailed in the previous Section, we calculate the PDF by considering the description for the EoS shown in Section~\ref{sec:Equationofstate}. As a first result, we display in Fig.~\ref{Corner_unif} the posterior PDFs for the empirical parameters using Uniform prior distributions related to Case~1 and Case~2.
\begin{figure*}
\centering
\includegraphics[scale=0.5, trim=0 0 0 21.7, clip=true]{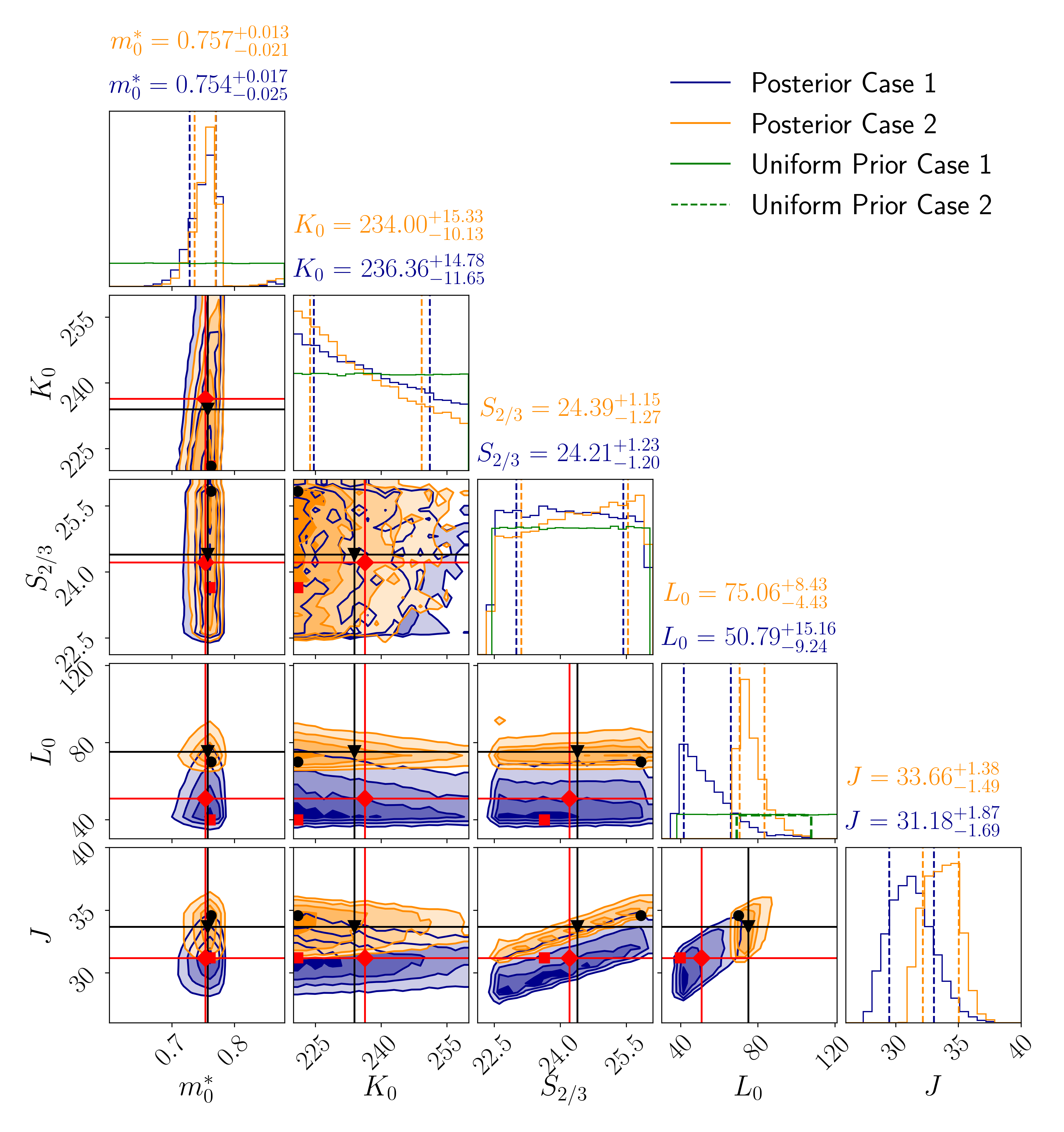}
\caption{Posterior distribution of the bulk parameters using Uniform prior for Case~1 (orange) and Case~2 (blue). The histograms shown on the diagonal represent the posterior marginal distribution of each parameter independently. The solid black (red) lines represent the median value of PDF for Case~1 (Case~2). The blue (orange) dashed lines indicate the values for $0.16$ and $0.84$ quantiles for Case~1 (Case~2). The black (red) dot (square) points refer to the median values or the Case~1 (Case~2). The black (red) triangle (diamond) is the most probable value or the Case~1 (Case~2), i.e., the modes. The $\sigma$ levels in the 2D histograms are $0.5\sigma$, $1\sigma$, $1.5\sigma$, and $2\sigma$. The prior limit is shown as a green line (dashed line) for Case~1 (Case~2) in the diagonal.}
\label{Corner_unif}
\end{figure*}
\begin{figure*}
\centering
\includegraphics[scale=0.5, trim=0 0 0 21.7,clip=true]{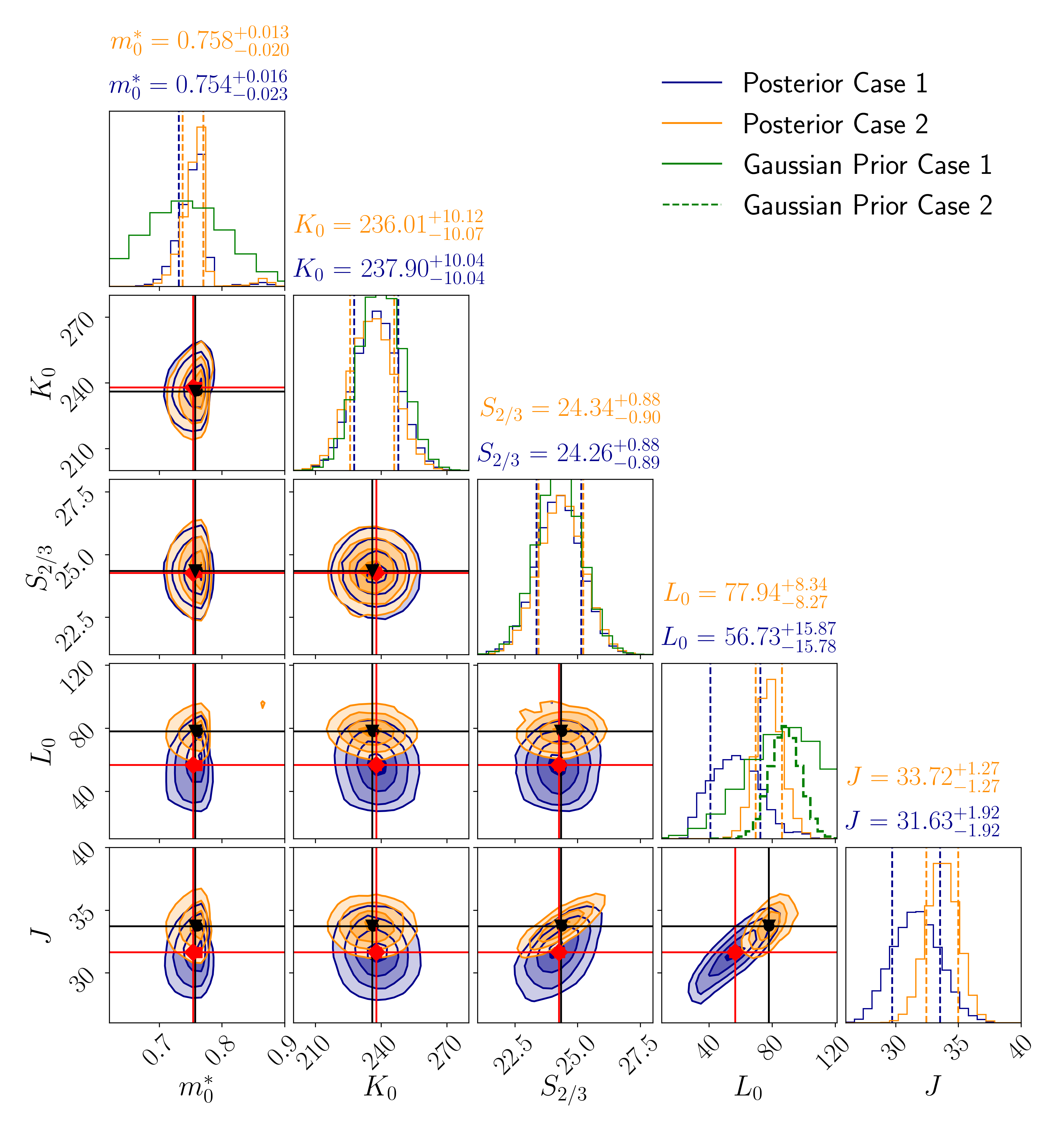}
\caption{Posterior distribution of the bulk parameters using a Gaussian prior. The same notation as in Fig.~\ref{Corner_unif}.}
\label{Corner_gauss}
\end{figure*}
For both cases, the effective mass shows a preference for larger values, presenting a very pronounced peak around the mean value. Notice that the histogram has a steep drop around $m_0^* = 0.8$. One of the reasons could be the upper limit of the interval chosen for the analysis ($0.88$), which is close to the most probable value. Both $K_0$ and $S_{2/3}$ present a flat behavior in its posterior. The incompressibility has a decreasing behavior and a trend of choosing lower values compared with the prior in both cases. Notice that for $S_{2/3}$, the final statistical analysis presents no abrupt changes in comparison with the uniform prior adopted for both cases. In particular, in Case~2 there is a general trend of increasing in the posterior in the direction of attaining the upper limit of the prior, namely, $\approx 26$~MeV. This is the number pointed out in~\cite{PREX-PRL126-172503} as being the correct value for $S_{2/3}$. The authors claim that $S_{2/3}\approx 26$~MeV since this quantity is tightly constrained by the binding energy of heavy nuclei. 
Concerning $L_0$, we can observe a preference for smaller values for both cases. Nonetheless, due to the upper limit being greater in the range of Case~2, it allows the slope of the symmetry energy to obtain values greater than those obtained in Case~1. 

After the calculation via Bayesian analysis, a 1D and 2D histograms were built for the values obtained for the symmetry energy in the saturation density, using the set of parameters obtained in the posterior distribution (see the last line of Fig.~\ref{Corner_unif}). Note that $J$ is distributed centered for both cases. This distribution is highly likely to be less than $\approx 34$~($\approx 36$)~MeV for Case~1 (Case~2) with a confidence interval (CI) of $90\%$. Although both distributions are centered, the one for Case~1 produces the most probable values around $31$~MeV against $34$~MeV for Case~2. 

By analyzing the posterior calculated using the Gaussian prior, Fig~\ref{Corner_gauss} one sees the same behavior for the effective mass as when the Uniform prior was used, that is, a sharp pick followed by a cut-off around $0.8$ for both cases studied. The posterior distributions for $K_0$ and $S_{2/3}$ do not differ from the prior. For these three quantities, Cases~1 e 2 have practically a similar comportment. The same does not occur for $L_0$, but the tendency to choose lower limit values is still the same. Both Cases have the posterior values shifted to the left in relation to their respective priors. For this quantity, the entire analysis with $90\%$ CI allows intervals around $33-84$~($64-93$)~MeV. The most probable value for the symmetry energy found after the analysis is practically the same obtained in Fig~\ref{Corner_unif}, for Case~1 (Case~2).

The boundaries for the bulk parameters along with other statistical quantities such as mean, mode, and median are shown in Table~\ref{table:sat-results} for the different priors used. 
\begin{table}
\setlength\tabcolsep{0.14cm}
\caption{Ranges for the bulk parameters obtained from the posterior by using the different priors. All quantities are in MeV, except $m_0^*$ (dimensionless). CI: confidence interval.}
\label{table:sat-results}
\begin{tabular}{lccccc} 
\hline
  & Mean & Mode & Median & $68\%$ CI & $90\%$ CI \\
\hline
\multicolumn{6}{c}{Uniform}\\
\hline
\multicolumn{6}{c}{Case 1}\\
$m_0^*$               & 0.751  & 0.761  & 0.754  & 0.729 -- 0.771     & 0.710 -- 0.777\\
$K_0$                 & 237.54 & 221.00 & 236.36 & 224.71 -- 251.14   & 221.39 -- 257.10\\
$\mathcal{S}_{2/3}$   & 24.22  & 23.65  & 24.21  & 23.01 -- 25.44     & 22.62 -- 25.84\\
$L_0$                 & 53.75  & 39.78  & 50.79  & 41.55 -- 65.95     & 39.05 -- 78.29\\
$J$                   & 31.28  & 31.18  & 31.18  & 29.49 -- 33.05     & 28.64 -- 34.37\\
\hline
\multicolumn{6}{c}{Case 2}\\ 
$m_0^*$               & 0.757  & 0.763  & 0.757  & 0.737 -- 0.770    & 0.720 -- 0.777\\ 
$K_0$                 & 235.88 & 221.00 & 234.00 & 223.87 -- 249.32  & 221.16 -- 256.22\\ 
$\mathcal{S}_{2/3}$   & 24.34  & 25.83  & 24.39  & 23.12 -- 25.54    & 22.66 -- 25.88\\
$L_0$                 & 77.00  & 69.97  & 75.06  & 70.62 -- 83.48    & 69.50 -- 91.38\\ 
$J$                   & 33.65  & 34.58  & 33.66  & 32.17 -- 35.04    & 31.52 -- 35.78\\ 
\hline
\multicolumn{6}{c}{Gaussian}\\
\hline
\multicolumn{6}{c}{Case 1}\\ 
$m_0^*$               & 0.753  & 0.760  & 0.754  & 0.731 -- 0.770     & 0.714 -- 0.777 \\
$K_0$                 & 237.91 & 237.88 & 237.90 & 227.87 -- 247.94   & 221.25 -- 254.43 \\
$\mathcal{S}_{2/3}$   & 24.25  & 24.28  & 24.26  & 23.37 -- 25.14     & 22.76 -- 25.73 \\
$L_0$                 & 58.01  & 56.32  & 56.73  & 40.96 -- 72.61     & 33.50 -- 83.82 \\
$J$                   & 31.68  & 31.61  & 31.63  & 29.71 -- 33.56     & 28.60 -- 34.81 \\
\hline
\multicolumn{6}{c}{Case 2}\\ 
$m_0^*$               & 0.758  & 0.761  & 0.758  & 0.737 -- 0.770    & 0.722 -- 0.777\\ 
$K_0$                 & 236.02 & 236.41 & 236.01 & 225.94 -- 246.12  & 219.49 -- 252.70\\ 
$\mathcal{S}_{2/3}$   & 24.34  & 24.36  & 24.34  & 23.44 -- 25.23    & 22.85 -- 25.81\\ 
$L_0$                 & 78.03  & 78.07  & 77.94  & 69.67 -- 86.29    & 64.20 -- 92.65\\ 
$J$                   & 33.72  & 33.73  & 33.72  & 32.46 -- 34.99    & 31.62 -- 35.80\\ 
\hline
\end{tabular}
\end{table}

To test the hypotheses raised by the used models, that is, for the priors and cases chosen, it is interesting to calculate the Bayes factor (BF) \citep{Jeffrey_book,Jeffreys_2009,Kohlinger_2019,Breschi2021}. The Bayes factor can be found by reason of the evidence of two models:
\begin{equation}
\mathcal{B}_{AB} = \frac{P(D|\mathcal{M}_A)}{P(D|\mathcal{M}_B)} = \frac{Z_A}{Z_B}.
\label{eq:BF}
\end{equation}
For the comparison, we define Case~1 as the set of parameters for $\mathcal{M}_A$. This assumption is made due to its less restrictive choice for the $L_0$ parameters. To interpret the BF we have used the suggestion proposed by \cite{Jeffrey_book} and summarized in \cite{Robert_1995} given in Table~\ref{table:jeffrey}. The results for the models used are shown in Table~\ref{table:BFUN1} and Table~\ref{table:BFGA1}.

\begin{table}
\setlength\tabcolsep{0.3cm}
\caption{Standard values provided by Jeffrey’s scale \citep{Jeffrey_book}, useful to compare two hypotheses by using the Bayes factor.}
\label{table:jeffrey}
\begin{tabular}{llc} 
\hline
 ${\rm log}_{10}(\mathcal{B}_{AB})$  & $\mathcal{B}_{AB}$ &  Evidence against $\mathcal{M}_A$ \\ 
\hline
  $<$ 0  &  $>$ 1     & Null hypothesis supported \\
0 to 0.5 & 1 to 3.2   & Not worth more than a bare mention \\
0.5 to 1 & 3.2 to 10  & Substantial \\
1 to 2   & 10 to 100  & Strong \\
$>$ 2    & $>$ 100    & Decisive \\
\hline
\end{tabular}
\end{table}
We can see a slight preference for the description that uses the uniform prior with the set of parameters defined in Case~2. This shows the preference for lower values of $L_0$, in agreement with recent studies that point out smaller ranges for this quantity. For example, we can quote the range of $L_0=(54\pm 8)$~MeV~\cite{Reinhard2021}. In this study, the authors determine the value of $L_0$ by analyzing the parity violation in $^{208}\rm Pb$. Still using the lead core, the reference~\cite{hu_2022} found the range of $L_0=(37-66)$~MeV using ab initio calculations. Recently, an analysis via Bayesian inference, taking into account the results of PREX-2 and CREX ~\cite{zhang_2022}, found $L_0=15.3^{+46.8}_{-41.5 }$~MeV. In the same vein, the authors reference larger values $L_0=(82.32\pm22.93)$~MeV~\citep{kumar_2023}.
\begin{table*}
\setlength\tabcolsep{0.5cm}
\caption{Evidence ($Z$) and Bayes factors, calculated through Eq.~\eqref{eq:BF}, for the models analysed. Model A refers to Case~1 with Uniform prior used.}
\label{table:BFUN1}
\begin{tabular}{lcccccc} 
\hline
Data  &  Prior & Model  & ln($Z$) &  $\mathcal{B}_{AB}$  &  ${\rm log}_{10}(\mathcal{B}_{AB})$ & 
Evidence against $\mathcal{M}_A$ on Jeffreys' scale \\ 
\hline
Case 1  & Uniform  & $\mathcal{M}_A$  & -94.06 & $-$  & $-$   & $-$ \\
Case 1  & Gaussian & $\mathcal{M}_B$  & -94.03 & 0.97 & -0.01 &  Negative (supports $\mathcal{M}_A$)  \\
Case 2  & Uniform   & $\mathcal{M}_B$  & -95.32 & 3.51 & 0.55  &  Substantial \\
Case 2  & Gaussian  & $\mathcal{M}_B$  & -95.09 & 2.80 & 0.45  &  Not worth more than a bare mention \\
\hline
\end{tabular}
\end{table*}

\begin{table*}
\setlength\tabcolsep{0.5cm}
\caption{The same as in Table~\ref{table:BFUN1} with Model A referring to Case~1 with Gaussian prior.}
\label{table:BFGA1}
\begin{tabular}{lcccccc} 
\hline
Data  &  Prior & Model  & ln($Z$) &  $\mathcal{B}_{AB}$  &  ${\rm log}_{10}(\mathcal{B}_{AB})$ & 
Evidence against $\mathcal{M}_A$ on Jeffreys' scale \\ 
\hline
Case 1  & Gaussian  & $\mathcal{M}_A$  & -94.03 & $-$  & $-$   & $-$ \\
Case 1  & Uniform   & $\mathcal{M}_B$  & -94.06 & 1.03 & 0.01  &  Not worth more than a bare mention \\
Case 2  & Uniform   & $\mathcal{M}_B$  & -95.32 & 3.62 & 0.56  &  Substantial \\
Case 2  & Gaussian  & $\mathcal{M}_B$  & -95.09 & 2.88 & 0.46  &  Not worth more than a bare mention \\
\hline
\end{tabular}
\end{table*}

\subsection{Symmetric matter, pure neutron matter, and neutron star properties}

At this point, we have all the probable values for the bulk parameters $m_0^*$, $K_0$, $\mathcal{S}_{2/3}$, $L_0$, and $J$. In order to verify how these numbers affect nuclear and stellar matter, we use the $68\%$ CI from the conditional probability distribution results. In addition, we also investigate the tidal deformabilities generated in the same way. The respective values of the coupling constants for these parameterizations are shown in Table~\ref{table:couplings}.

In the case of symmetric nuclear matter, we analyze the obtained parametrizations in the pressure versus density plot, see Fig.~\ref{fig_pxratio-y05-flow_EoS}. 
\begin{figure}
\centering
\includegraphics[width=\columnwidth]{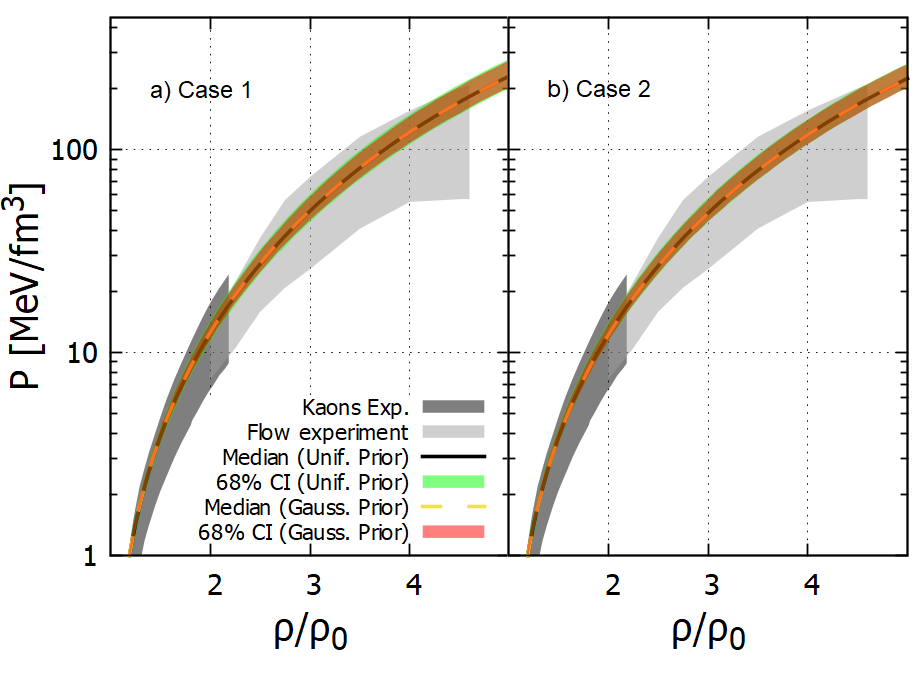}
\caption{Pressure versus $\rho/\rho_0$ (symmetric matter) for the parametrizations obtained from the Bayesian analysis. The flow constraint (light gray area) was taken from~\citep{pawel2002}. The dark gray band was extracted from~\citep{LYNCH2009427}. Green and red bands correspond to the parametrizations at $68\%$ CI (posteriors) using Uniform and Gaussian priors, respectively. Full (Uniform prior used) and dashed (Gaussian prior used) curves represent those parametrizations related to the median values of the bulk parameters analyzed.}
\label{fig_pxratio-y05-flow_EoS}
\end{figure}
In these figures, the shaded areas were obtained from analysis of experimental data from nuclear collision~\citep{pawel2002}, and from the investigation of kaon production in heavy ion collisions\citep{LYNCH2009427}, respectively. We can see that the resulting EoSs satisfies the two constraint regions mentioned above. Furthermore, the description of the EoSs generated by using the results of the posterior, with the Uniform and Gaussian priors, is practically identical within $68\%$ CI, which is evidenced by the orange region, where the description generated by each of the priors overlaps. The main difference is related to the median values, which produce an EoS closer to the edge for Case~2 when compared to Case~1.

The investigation of pure neutron matter is also important because it is a good approximation of the constituent matter in NS. In \cite{pawel2002}, the authors analyzed the data of the flow of particles ejected from collisions between \mbox{$^{197}$Au} ions. The authors proposed a constraint for pressure depending on density through extrapolation of the data, including asymmetry terms with strong (\textit{stiff}) or weak (\textit{soft}) dependence of density on pressure. The delimited region found can be seen in Fig.~\ref{fig_pxratio-y0-Asy_EoS} where, once again, it is verified that the resulting EoSs satisfy the constraint.
\begin{figure}
\centering
\includegraphics[width=\columnwidth]{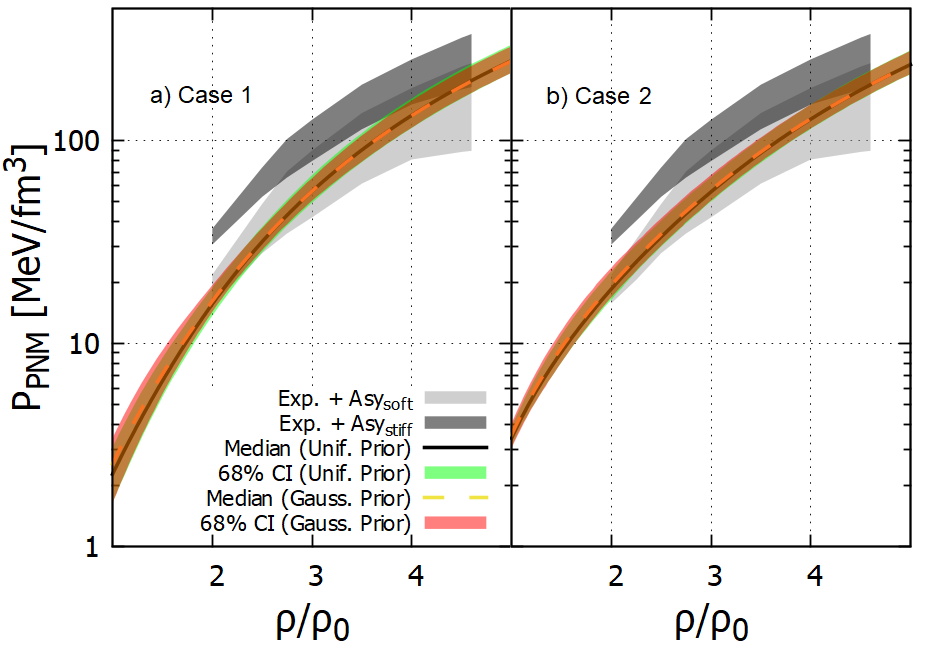}
\caption{Pressure versus $\rho/\rho_0$ (pure neutron matter) for the parametrizations obtained from the Bayesian analysis. The gray bands were taken from \citep{pawel2002}. Green and red bands correspond to the parametrizations at $68\%$ CI (posteriors) using Uniform and Gaussian priors, respectively. Full (Uniform prior used) and dashed (Gaussian prior used) curves represent those parametrizations related to the median values of the bulk parameters analyzed.}
\label{fig_pxratio-y0-Asy_EoS}
\end{figure}

Still, within the scope of PNM, microscopic studies based on chiral effective field theory (chEFT) have been done over the years, see \citep{Hebeler_2013,Kruger_2013,Drischler_2016}, for instance. In these works, low-moment expansions are used to describe the nuclear force. Interactions between nucleons are taken as point-like or via pion exchange. The corresponding parameters are adjusted through the observable of two and three bodies. In Fig.~\ref{fig:E_vs_rho_PNM} it is displayed a comparison of our results, i.e., the parametrizations with $68\%$ CI from the conditional probability distribution confronted with the bands in the energy per particle versus density plot related to chEFT calculations.
\begin{figure}
\centering
\includegraphics[width=\columnwidth]{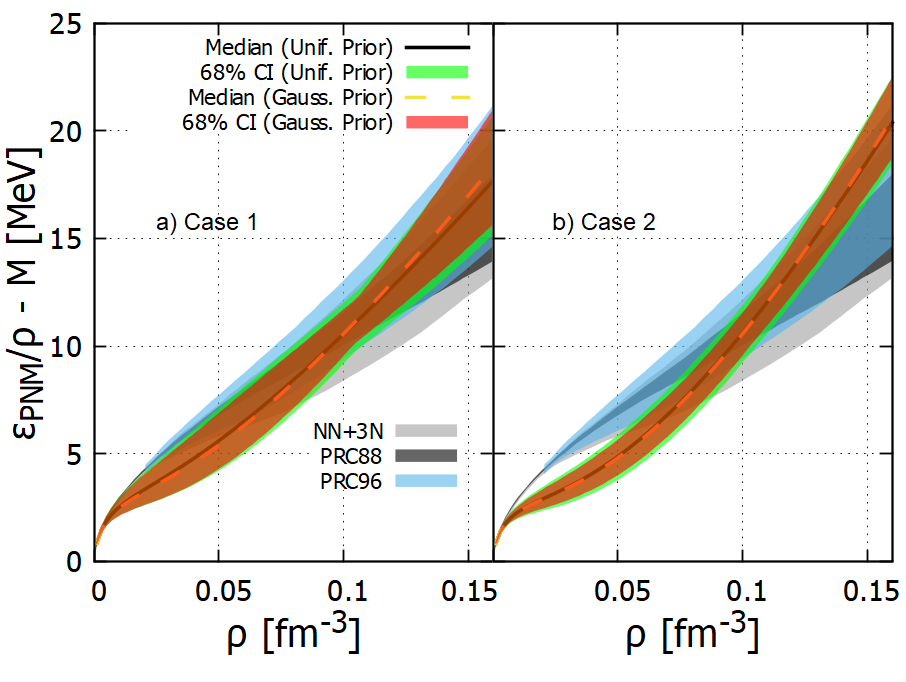}
\caption{Energy per particle of pure neutron matter as a function of density. The grey, dark-grey, and light-blue bands, derived from chEFT, were extracted from \citep{Hebeler_2013,Kruger_2013,Drischler_2016}. Green and red bands correspond to the parametrizations at $68\%$ CI (posteriors) using Uniform and Gaussian priors, respectively. Full (Uniform prior used) and dashed (Gaussian prior used) curves represent those parametrizations related to the median values of the bulk parameters analyzed.}
\label{fig:E_vs_rho_PNM}
\end{figure}
We see that the parametrizations of Case~2 present greater deviations, especially for the lower-density region. 

In the context of stellar matter, we investigated how the results corresponding to the EoS parameters up to $68\%$ CI affect this system. The results for the mass-radius diagrams are shown in Fig~\ref{MxR_curves}. 
\begin{figure}
\centering
\includegraphics[width=\columnwidth, trim=0 0 0 0, clip=true]{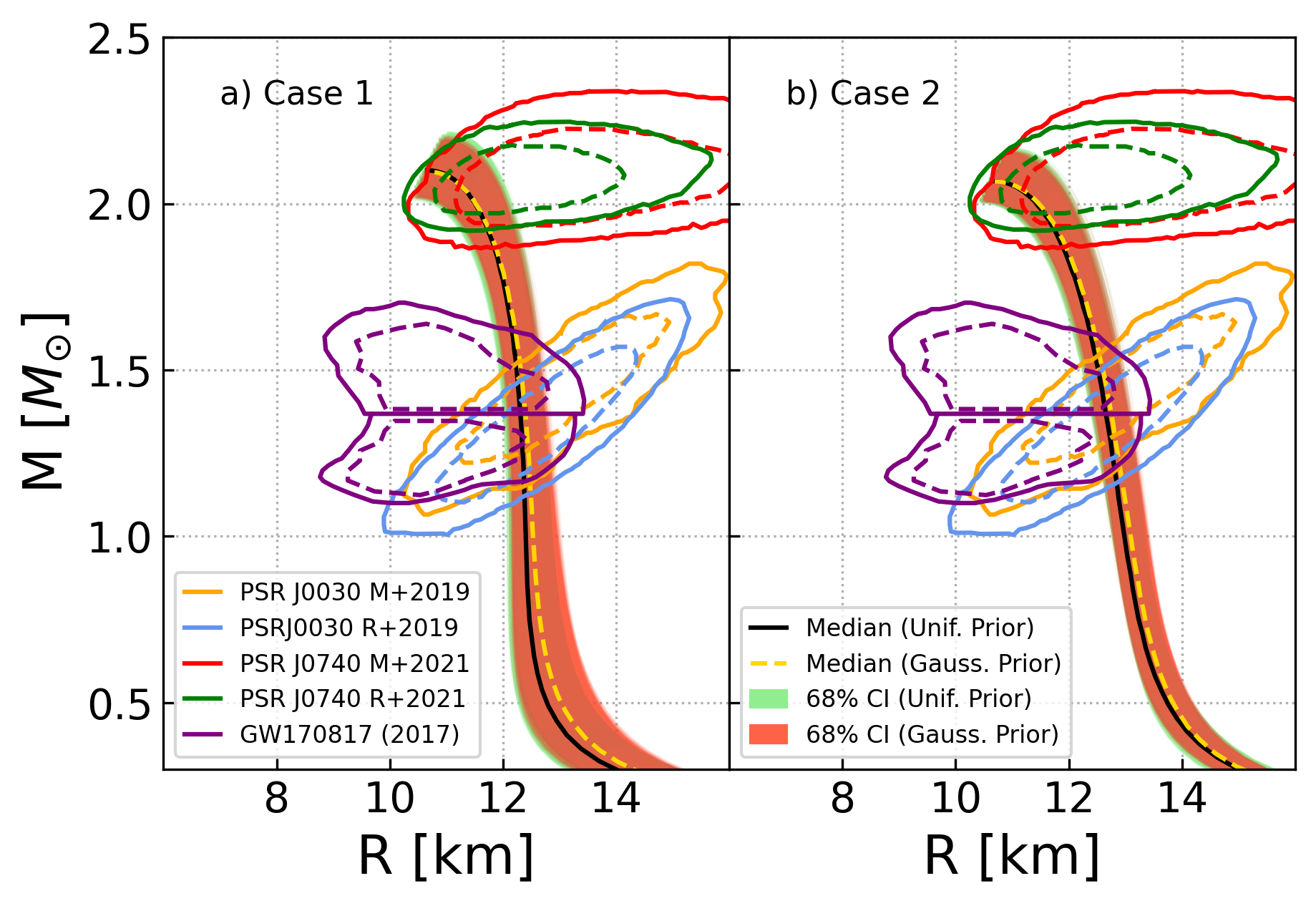}
\caption{Mass-radius diagrams for the parametrizations obtained from the Bayesian analysis. Green and red bands correspond to the parametrizations at $68\%$ CI (posteriors) using Uniform and Gaussian priors, respectively. Full (Uniform prior used) and dashed (Gaussian prior used) curves represent those parametrizations related to the median values of the bulk parameters analyzed. The contours for the mass-radius constrained by the NICER mission are also shown, namely, orange~\citep{Miller_2019} and blue~\citep{Riley_2019} ones PSR~J0030+0451; red~\citep{Miller_2021} and green~\citep{Riley_2021} ones for PSR~J0740+6620} The violet contour represents the GW170817 event \citep{Abbott2018}.
\label{MxR_curves}
\end{figure}
In this figure, we also show the contours corresponding to the NICER~\citep{Miller_2019,Miller_2021,Riley_2019,Riley_2021} and LVC~\citep{Abbott_2017,Abbott2018} data. We can observe that Case~1 and Case~2 are very similar to each other, especially for $M/M_\odot>1$. The inferred mass-radius relationship is also shown in Figs.~\ref{MxR_busters} and \ref{MxR_qLM} in relation to the other observables used in this article~\citep{ozel2016dense,Natalia2017}. Note that the results also support a CI of $68\%$ for most of the sources used in the work. However, the source described by $\omega$~Cen in Fig.~\ref{MxR_qLM} has a small distance from the results obtained for $68\%$. This distance is small for Case~1 in comparison with Case~2. This is due to the values of $L_0$ presented by the final ranges as results of the Bayesian analysis, as one can see in Table~\ref{table:sat-results}. Notice that the smaller range for $L_0$ is the one given by Case~1. For this case, the radii of the stars generated are also smaller in comparison with the ones obtained in Case~2, see Figs.~\ref{MxR_curves} to~\ref{MxR_qLM}. The effect of reduction of $R$ as $L_0$ decreases was also verified in~\cite{Fattoyev2013,Alam_2016,Zhang_2019,PRC101,dmnosso2}.
\begin{figure}
\centering
\includegraphics[width=\columnwidth, trim=0 0 0 0, clip=true]{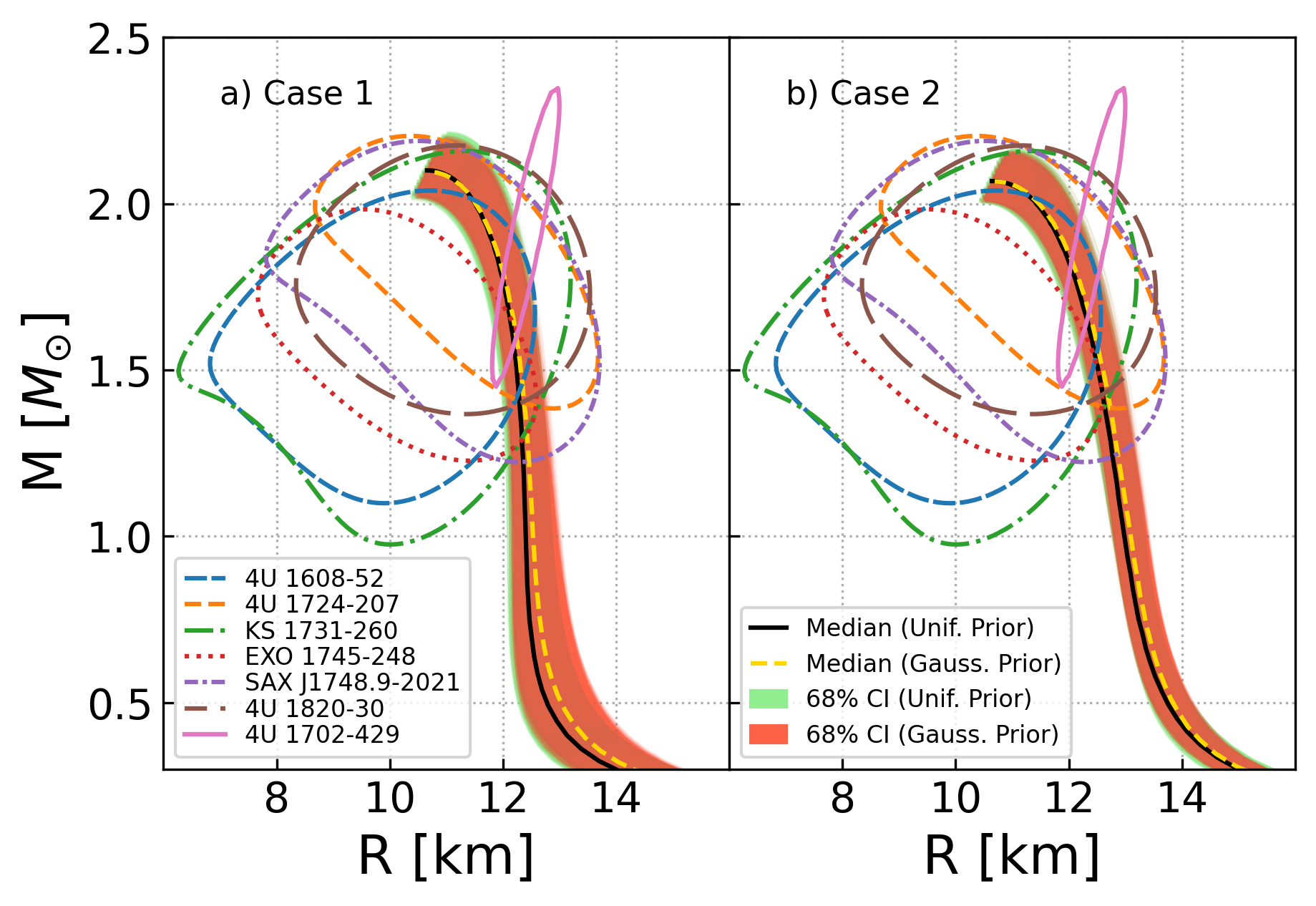}
\caption{Same as Fig.~\ref{MxR_curves}, but now compared with $7$ more observables used in the paper. 4U 1702-429~\citep{Natalia2017}. For the remaining ones see \citep{ozel2016dense} and the references therein.}
\label{MxR_busters}
\end{figure}
\begin{figure}
\centering
\includegraphics[width=\columnwidth, trim=0 0 0 0, clip=true]{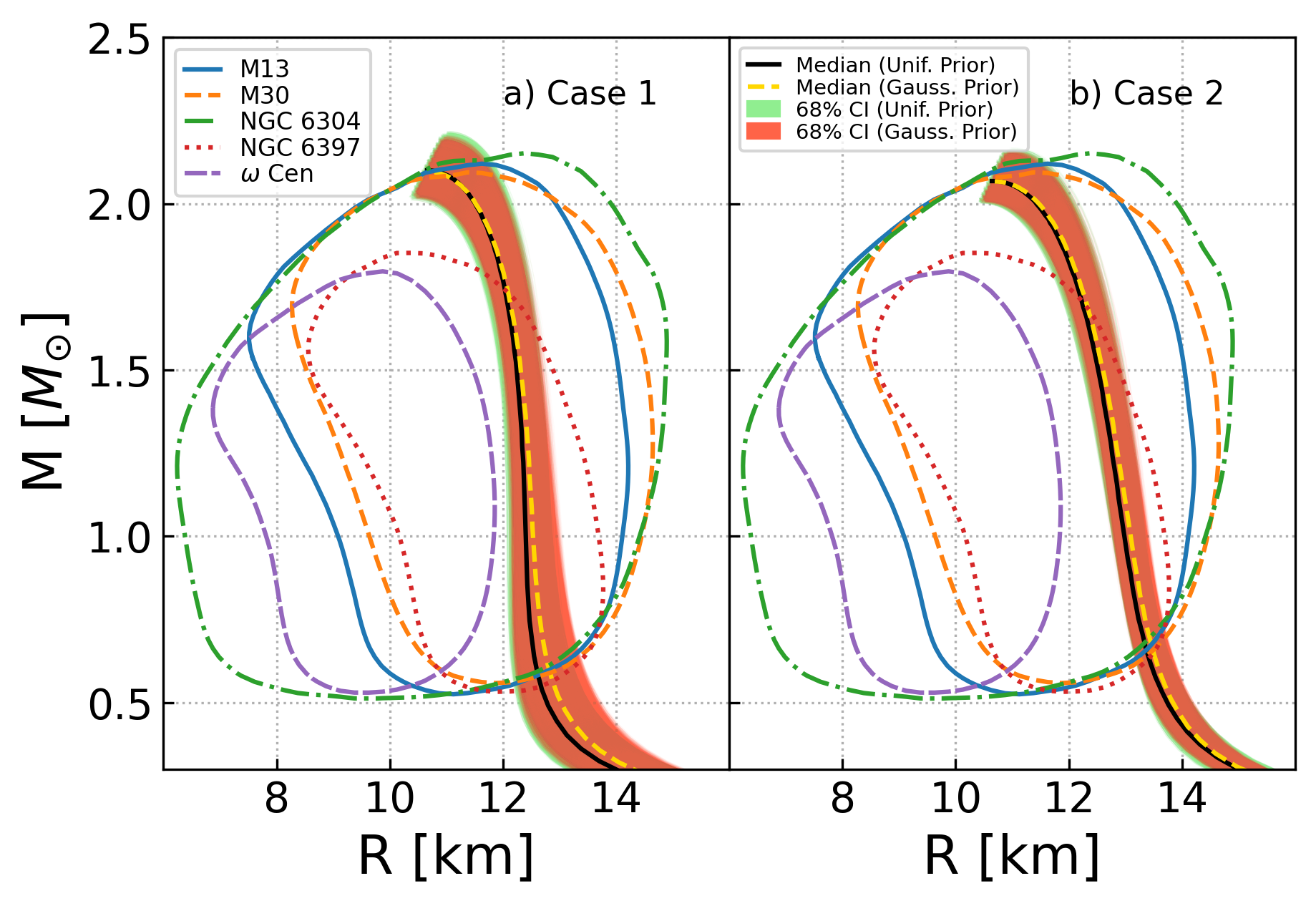}
\caption{Same as Fig.~\ref{MxR_curves}, but now compared to the remaining $5$ observables. See \citep{ozel2016dense} and the references therein.}
\label{MxR_qLM}
\end{figure}
\begin{table}
\setlength\tabcolsep{0.14cm}
\caption{Some macro properties resulting from the equations of state within $68\%$ CI. See 4th column in Table~\ref{table:sat-results}. }
\label{table:MmaxRmax}
\begin{tabular}{ccccc} 
\hline
& $M_{\rm max}$ $(M_{\odot})$ & $R_{\rm max}$ (km) & $R_{1.4}$ (km) & $\Lambda_{1.4}$\\
\hline
\multicolumn{5}{c}{Uniform}\\
\hline
Case 1 & 2.01 -- 2.22 & 10.39 -- 11.06 & 11.97 -- 12.73 & 354.05 -- 474.02\\
Case 2 & 2.00 -- 2.17 & 10.41 -- 10.98 & 12.34 -- 13.06 & 378.52 -- 505.15\\
\hline
\multicolumn{5}{c}{Gaussian}\\
\hline
Case 1 & 2.00 -- 2.20 & 10.43  -- 11.02 & 12.04 -- 12.84 &  357.68 -- 481.32\\
Case 2 & 2.00 -- 2.16 & 10.53 -- 10.96 & 12.42 -- 13.12 & 383.76 -- 541.65\\
\hline
\end{tabular}
\end{table}
In Table~\ref{table:MmaxRmax} the values for the maximum properties for each prior with $68\%$ CI are shown. The ranges of these values for the macro properties of the NS, namely, maximum mass ($M_{\rm max}$), its respective radius ($R_{\rm max}$), the radius of the $1.4M_\odot$ neutron star ($R_{1.4}$), and dimensionless tidal deformability of the $1.4M_\odot$ neutron star ($\Lambda_{1.4}$) are practically the same for Uniform and Gaussian priors used in each case analyzed. 

Recently, the authors of \cite{Breschi2021} have collected recent estimates for $R_{1.4}$. We display this band in Fig.~\ref{radius1.4}, as well as another prediction for this quantity found in~\cite{huth2022}, also from the Bayesian inference. Since all of them are estimations based on a statistical analysis taking into account $90\%$ or more confidence intervals, we compare our results using also $90\%$ CI. Notice that all points obtained in our work (circles) agree with these estimations, showing a trend toward values closer to the upper limits.
\begin{figure}
\centering
\includegraphics[width=\columnwidth]{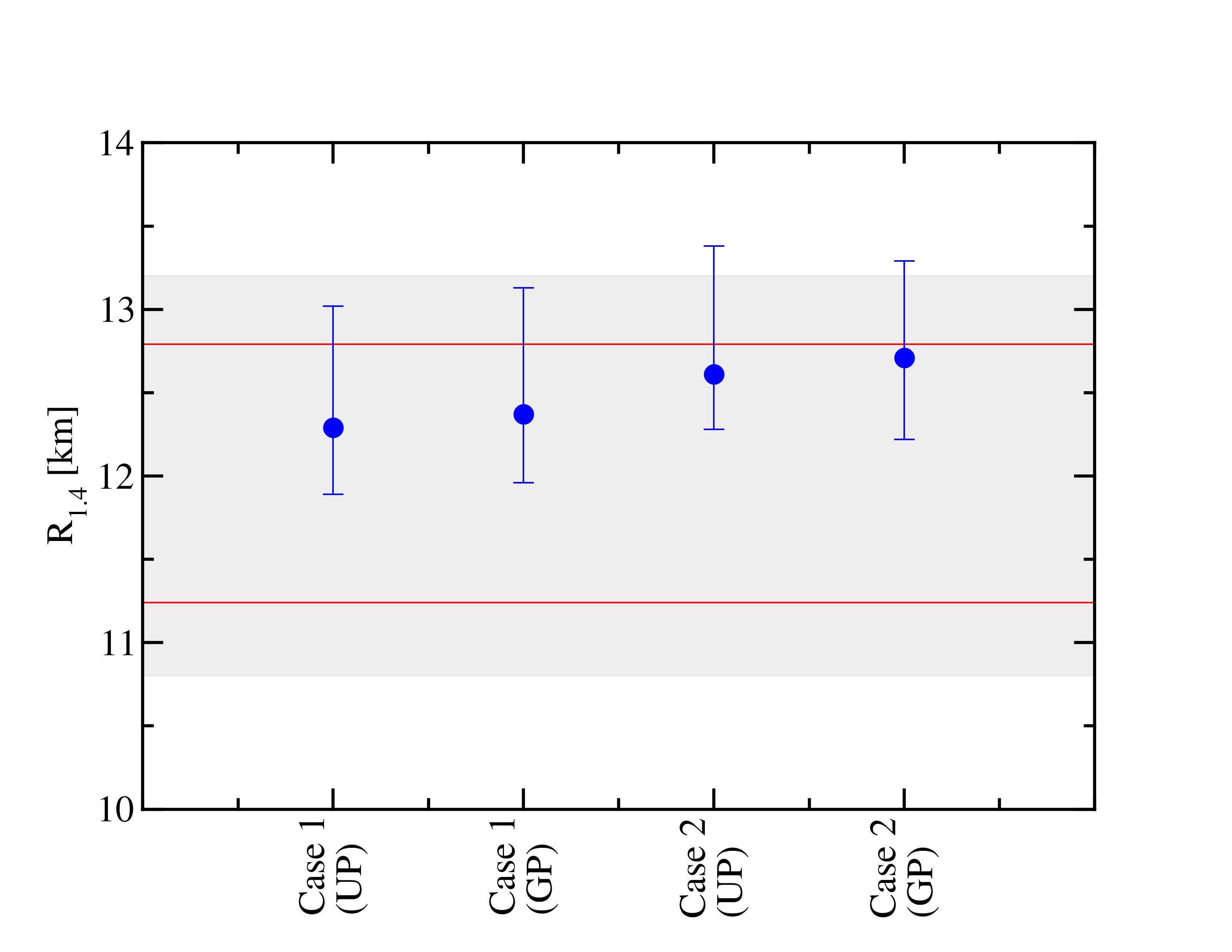}
\caption{Predictions for $R_{1.4}$ calculated from the parametrizations obtained from the Bayesian analysis (circles). In this case, we show the results for the NS radius for $1.4M_{{\rm sun}}$ at $90\%$ CI (posteriors), when the Uniform (UP) and Gaussian (GP) priors are used}. Gray band extracted from \citep{Breschi2021}, and red lines taken from \citep{huth2022}.
\label{radius1.4}
\end{figure}

Finally, we calculate the dimensionless tidal deformability and depict how this quantity depends on $M$ in Fig.~\ref{lambda_x_m}. 
\begin{figure}
\centering
\includegraphics[width=\columnwidth]{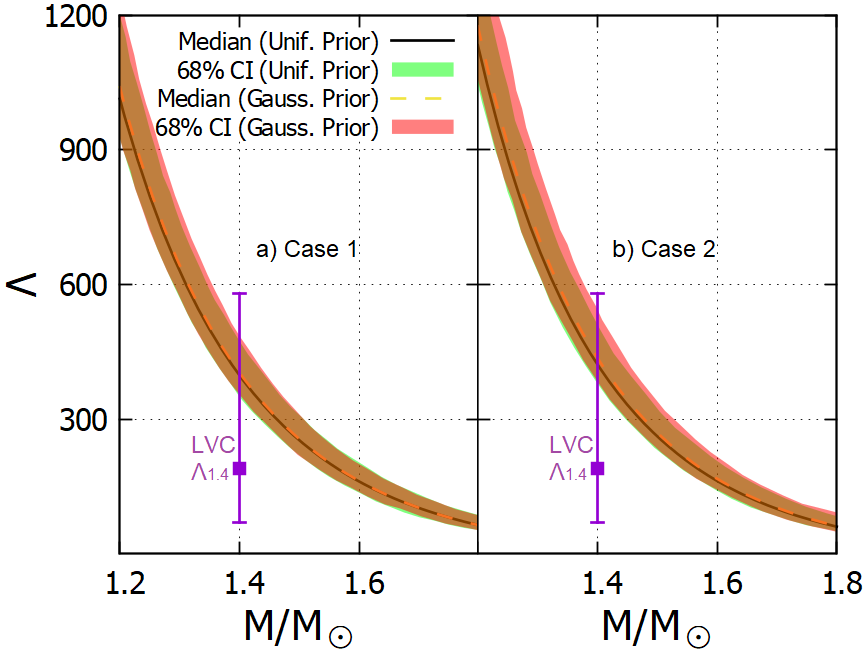}
\caption{$\Lambda$ as a function of $M$ for the parametrizations obtained from the Bayesian analysis. Green and red bands correspond to the parametrizations at $68\%$ CI (posteriors) using Uniform and Gaussian priors, respectively. Full (Uniform prior used) and dashed (Gaussian prior used) curves represent those parametrizations related to the median values of the bulk parameters analyzed. The purple square is the result of $\Lambda_{1.4} = 190^{+390}_{-120}$ obtained by LVC \citep{Abbott2018}.}
\label{lambda_x_m}
\end{figure}
The description for both priors is almost identical, and we can also see that all parametrizations produce the $\Lambda_{1.4}$ inside the range predicted by LVC regardless of the case used in the Bayesian analysis. When compared to Case~1, Case~2 has slightly greater values for $\Lambda$, especially at lower values of $M$. For $\Lambda_{1.4}$ specifically, some studies have already shown that this quantity is sensitive to $L_0$~\citep{Fattoyev2013,Alam_2016,Zhang_2019,PRC101,dmnosso2}. In our study, Case~2 has greater values for $L_0$ and as a consequence, $\Lambda$ is greater in general. In Fig.~\ref{lambda_x_lambda} we show the dimensionless tidal deformability for each of the stars of the binary system related to the GW170817 event, with component masses $M_1$, in the range of $1.37\leqslant M_1/M_\odot \leqslant 1.60$~\citep{Abbott_2017}. The mass of the companion star is calculated through the relationship between $M_1$, $M_2$, and the chirp mass, namely, $\mathcal{M} = (M_1M_2)^{3/5}/(M_1+M_2)^{1/5}=1.188M_\odot$~\citep{Abbott_2017}. 
\begin{figure}
\centering
\includegraphics[width=\columnwidth]{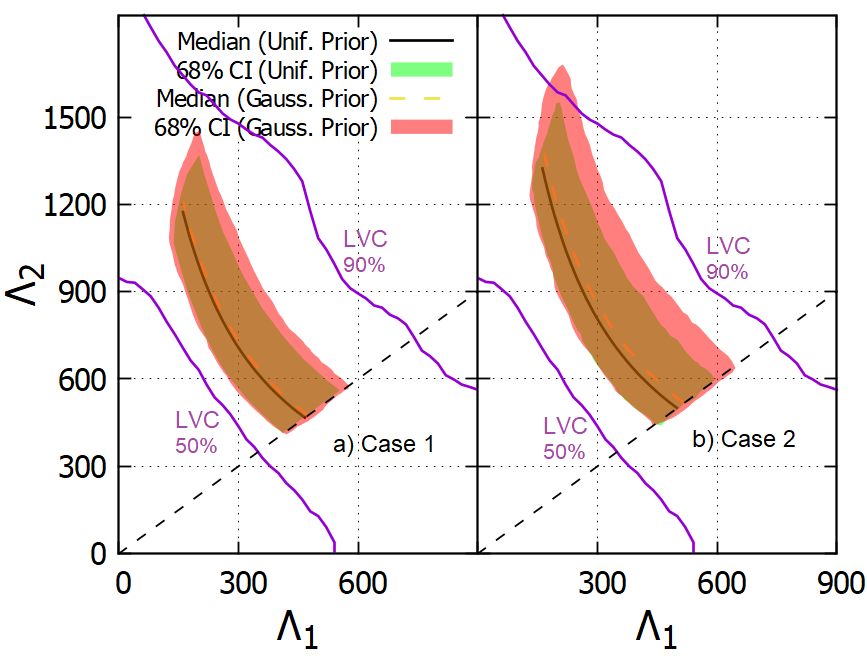}
\caption{Dimensionless tidal deformabilities for the case of high-mass ($\Lambda_1$) and low-mass ($\Lambda_2$) components of the GW170817 event. The purple lines are the results obtained by LVC \citep{Abbott2018} at $50\%$ and $90\%$ CI. The diagonal dashed lines correspond to $\Lambda_1=\Lambda_2$. Green and red bands correspond to the parametrizations at $68\%$ CI (posteriors) using Uniform and Gaussian priors, respectively. Full (Uniform prior used) and dashed (Gaussian prior used) curves represent those parametrizations related to the median values of the bulk parameters analyzed.}
\label{lambda_x_lambda}
\end{figure}
The deformabilities are computed using the resulting EoS within $68\%$ CI when both priors are used. We also show contour lines at the $50\%$ and $90\%$ confidence intervals (purple lines) associated with the GW170817 event. We can notice that all the results obtained in this work are inside the LVC predictions.

\section{Summary and Final Remarks}
\label{sec:Final}

In the paper, we have performed a Bayesian analysis in order to determine ranges for the bulk parameters of nuclear matter, namely, $m_0^*$ (effective mass ratio at $\rho=\rho_0$), $K_0$ (incompressibility at $\rho=\rho_0$), $\mathcal{S}_{2/3}$ (symmetry energy at $\rho=2\rho_0/3$), and $L_0$ (symmetry energy slope at $\rho=\rho_0$). For the Likelihood functions used as input for the Bayesian method, we have used 16 different sources of masses and radii, where 11 of them were extracted from~\cite{ozel2016dense} (6 are thermonuclear busters called 4U 1820–30, SAX J1748.9–2021, EXO 1745–248, KS 1731–260, 4U 1724–207, and 4U 1608 –52. 5 were obtained from the Quiescent Low-Mass X-ray Binaries M13, M30, NGC 6304, NGC 6397, and $\omega$ Cen), 1 taken from the neutron star X-ray analysis in 4U 1702–429~\citep{Natalia2017}, 2 components}from the LVC~\citep{samples-ligo}, and 2 from mass-radius measurements provided by the NICER mission~\citep{Riley_2019, Riley_2021}. From the ranges obtained as output, we generated the corresponding optimal parametrizations of the relativistic mean-field model used here, the one that takes into account the $\omega-\rho$ interaction in its structure. The inclusion of this term allowed us to control not only the symmetry energy but also its slope. Consequently, we were able to introduce specific values for $L_0$ as input in our analysis, namely, the ones inferred from the recent PREX-II experimental data~\citep{PREX-ESSICK,PREX-PRL126,PREX-PRL126-172503,PREX-PRR4-L022054}. 

We split our analysis into two sets named Case~1 and Case~2, both of them with two different priors, Uniform and Gaussian distributions. These sets take into account the extremes and intersections of the $L_0$ values obtained through the analysis of the PREX-II results aforementioned. The results show that the output values for $L_0$, the ones found by the Bayesian analysis, are impacted by the choice of the distribution, more specifically, the most probable value for this quantity (mode) and its lower limit for the $68\%$ and $90\%$ confidence intervals for the Case~1. This occurs because the Gaussian distribution allows possible choices of lower $L_0$ values. The same is not true for the Uniform distribution, in which lower values for $L_0$ have zero probability. Our posterior results also show that for a major (minor) restriction in the input values for $L_0$ in the Bayesian analysis for both prior distributions, Uniform and Gaussian, Case 1 (Case 2), the output results for this quantity change considerably. For example, the posterior values using the Uniform prior are $L_0 = 50.79^{+15.16}_{-9.24}$~MeV ($L_0 = 75.06^{+8.43}_{-4.43}$~MeV) for $68\%$ CI. These values are related to symmetry energy in the range of $J = 31.18^{+1.87}_{-1.69}$~MeV ($J = 33.66^{+1.38}_{-1.49}$~MeV), all of them compatible with values recently predicted in the literature~\citep{Adhikari2021,Adhikari2022,PREX-PRL126-172503,Brett2022}. 

Finally, we have shown that the EoS with $68\%$ CI obtained from the Bayesian analysis describes satisfactorily well both symmetric nuclear matter and pure neutron matter at sub- and supra-saturation density regimes. In the stellar matter, our findings show that the parameterizations produce NS with a maximum mass of $M_{\rm max} = (2.115\pm0.105)M_\odot$ for Case~1 and $M_{\rm max} = (2.085\pm0.085)M_\odot$ for Case~2, both for posterior distribution when the Uniform prior is used. The corresponding radii related to these masses are $R_{\rm max} = (10.725\pm0.335)$~km and $R_{\rm max} = (10.695\pm0.285)$~km, respectively. Furthermore, the values found for the radius of the $1.4M_\odot$ neutron star is $R_{1.4} = (12.35\pm0.38)$~km for Case~1, and $R_{1.4} = (12.7\pm0.36)$~km for Case~2. These values are fully compatible with those recently obtained in~\cite{Breschi2021} and ~\cite{huth2022}. The numbers predicted by the optimal parametrizations regarding the dimensionless tidal deformabilities $\Lambda_1$, and $\Lambda_2$ (binary neutron star system), and $\Lambda_{1.4}$ (related to the $1.4M_\odot$ neutron star) also agrees with the observational constraints imposed by LVC from the analysis of the GW170817 event, namely, detection of gravitational waves from the merger of two neutron stars.

\section*{Acknowledgements}
This work is a part of the project INCT-FNA Proc. No. 464898/2014-5, and it was financed in part by the Coordena\c c\~ao de Aperfei\c coamento de Pessoal de N\'ivel Superior – Brasil (CAPES) – Finance Code 001 - (B.A.M.S., M.D.) . It is also supported by Conselho Nacional de Desenvolvimento Cient\'ifico e Tecnol\'ogico (CNPq) under Grants 433369/2018-3, and 308528/2021-2 (M.D.) and 312410/2020-4 (O.L.). We also acknowledge Fundação de Amparo \`a Pesquisa do Estado de S\~ao Paulo (FAPESP) under Thematic Project No. 2017/05660-0 (M. D., C. H. L., and O. L.), Grant No. 2020/05238-9 (M.D., C.H.L., and O.L.) and Project 2022/03575-3 (O.L). The authors also thank LAB-CCAM from ITA for computational support. The authors acknowledge POWER OF DATA TECNOLOGIA LTDA for providing a technological environment (big data \& analytics) for data processing.

\section*{DATA AVAILABILITY STATEMENT}
This manuscript has no associated data or the data will not be deposited. All data generated during this study are contained in this published article.

\bibliographystyle{mnras.bst}
\bibliography{references-revised}{}

\appendix
\section{EoS coupling constants}

In Table~\ref{table:couplings} we furnish the coupling constants for the parametrizations that produce the nuclear bulk parameters presented in Table~\ref{table:sat-results}.
\begin{table}
\setlength\tabcolsep{0.15cm}
\caption{Coupling constants related to the parametrizations in which the bulk parameters are shown in Table~\ref{table:sat-results}.}
\label{table:couplings}
\begin{tabular}{lcccccc} 
\hline
  &  $g_{\sigma}^2/m_{\sigma}^2$ & $g_{\omega}^2/m_{\omega}^2$ & $g_{\rho}^2/m_{\rho}^2$ & $100A/g_{\sigma}^3$ & $B/g_{\sigma}^4$ & $10\alpha$ \\
  & (fm$^2$) & (fm$^2$) & (fm$^2$) & (fm$^{-1}$) & ($10^{-3}$) & \\
\hline
\multicolumn{7}{c}{Uniform}\\
\hline
\multicolumn{7}{c}{Case 1}\\
Mean & 10.94 & 5.81 & 6.85 & -3.13 & -4.56 & 1.00 \\
Mode & 10.84 & 5.51 & 7.47 & -3.86 & -6.85 & 1.73 \\
Median & 10.88 & 5.74 & 7.00 & -3.23 & -4.65 & 1.12 \\
$68\%$ L.L. & 11.76 & 6.48 & 7.12 & -2.83 & -5.93 & 1.41 \\
$68\%$ U.L. & 10.18 & 5.24 & 6.99 & -3.23 & -0.80 & 0.75 \\
$90\%$ L.L. & 12.34 & 7.03 & 7.39 & -2.47 & -5.69 & 1.49 \\
$90\%$ U.L. & 9.90 & 5.04 & 6.84 & -3.15 & 1.68 & 0.49 \\
\hline
\multicolumn{7}{c}{Case 2}\\ 
Mean & 10.79 & 5.65 & 6.09 & -3.32 & -4.62 & 0.42 \\
Mode & 10.8 & 5.48 & 7.07 & -3.90 & -6.85 & 0.62 \\
Median & 10.80 & 5.63 & 6.16 & -3.38 & -4.89 & 0.46 \\
$68\%$ L.L. & 11.54 & 6.25 & 5.61 & -3.03 & -6.17 & 0.48 \\
$68\%$ U.L. & 10.22 & 5.25 & 6.55 & -3.28 & -1.22 & 0.36 \\
$90\%$ L.L. & 12.06 & 6.74 & 5.39 & -2.68 & -5.99 & 0.47 \\
$90\%$ U.L. & 9.92 & 5.06 & 6.56 & -3.18 & 1.37 & 0.23 \\
\hline
\multicolumn{7}{c}{Gaussian}\\
\hline
\multicolumn{7}{c}{Case 1}\\ 
Mean & 10.90 & 5.77 & 6.67 & -3.16 & -4.47 & 0.88 \\
Mode & 10.67 & 5.54 & 6.76 & -3.36 & -4.17 & 0.97 \\
Median & 10.85 & 5.72 & 6.73 & -3.20 & -4.42 & 0.92 \\
$68\%$ L.L. & 11.66 & 6.41 & 7.37 & -2.83 & -5.68 & 1.42 \\
$68\%$ U.L. & 10.23 & 5.24 & 6.63 & -3.33 & -1.46 & 0.58 \\
$90\%$ L.L. & 12.23 & 6.92 & 8.46 & -2.55 & -5.81 & 1.84 \\
$90\%$ U.L. & 9.93 & 5.05 & 6.65 & -3.25 & 1.04 & 0.37 \\
\hline
\multicolumn{7}{c}{Case 2}\\ 
Mean & 10.76 & 5.62 & 6.06 & -3.35 & -4.57 & 0.40 \\
Mode & 10.67 & 5.53 & 6.08 & -3.42 & -4.38 & 0.40 \\
Median & 10.77 & 5.62 & 6.06 & -3.34 & -4.58 & 0.40 \\
$68\%$ L.L. & 11.49 & 6.22 & 5.80 & -3.02 & -5.98 & 0.51 \\
$68\%$ U.L. & 10.25 & 5.24 & 6.33 & -3.39 & -1.83 & 0.29 \\
$90\%$ L.L. & 12.03 & 6.69 & 5.67 & -2.74 & -6.17 & 0.60 \\
$90\%$ U.L. & 9.94 & 5.04 & 6.50 & -3.32 & 0.71 & 0.21 \\
\hline
\end{tabular}
\end{table}

\bsp	
\label{lastpage}
\end{document}